\documentstyle[preprint,aps,prbbib]{revtex}
\tightenlines
\begin{document}
\draft
\title{Critical chain length and superconductivity emergence
in oxygen-equalized pairs of YBa$_2$Cu$_3$O$_{6.30}$}
\author{P. Manca \and S. Sanna}
 \address{Dipartimento di Fisica and Istituto Nazionale di Fisica della Materia
 (INFM),
Universit\`{a} di Cagliari,\\
 Cittadella Universitaria, 09042 Monserrato (CA), Italy.}
\author{G. Calestani}
\address{Dipartimento di Chimica Generale ed Inorganica, Chimica Analitica e
Chimica Fisica, Universit\`{a} di Parma,
 \\ Viale delle Scienze, 43100 Parma, Italy.}
\author{A. Migliori}
\address{CNR-LAMEL, Area della Ricerca di Bologna, \\ Via Gobetti 101,
Bologna, Italy.}
\author{R. De Renzi \and G. Allodi}
\address{Dipartimento di Fisica and Istituto Nazionale di Fisica della Materia
(INFM), Universit\`{a} di Parma, \\ Viale delle Scienze 7a, 43100 Parma,
 Italy.}

\date{\today}
\maketitle
\begin{abstract}
The oxygen-order dependent emergence of superconductivity in
YBa$_2$Cu$_3$O$_{6+x}$ is studied, for the first time in a
comparative way, on pair samples having the same oxygen content
and thermal history, but different Cu(1)O$_x$ chain arrangements
deriving from their intercalated and deintercalated nature. Structural
and electronic non-equivalence of pairs samples is detected in the
critical region and found to be related, on microscopic scale, to
a different average chain length, which, on being experimentally
determined by nuclear quadrupole resonance (NQR), sheds new light
on the concept of critical chain length for hole doping
efficiency.
\end{abstract}
The peculiarity of  YBa$_2$Cu$_3$O$_{6+x}$(123), the
superconductor that still plays an important role in ongoing
efforts to elucidate the mechanism of high-T$_c$
superconductivity, is the existence of a charge reservoir, the
...-Cu-O-Cu-.. chain system in the Cu(1)O$_x$  plane, far
removed from the superconducting Cu(2)O$_2$ sheets. Its structural
order drives the whole crystal structure in a variety of
superstructures which have been 
observed\cite{Alario,Reyes,3,Stratilov,5,6,7,Schleger} and
modeled theoretically\cite{Wille,10,deFont,Ceder,Poulsen} in the whole
compositional $x$-range. Beside the tetragonal ({\bf T},
empty chain) and orthorhombic-I ({\bf OI}, full chain)
structures that characterize the end members of the compositional
$x$-range  ($x$=0 and 1 respectively), at least two orthorhombic
modifications, ortho-II ({\bf OII}) and ortho-III
({\bf OIII}), occurring around the ideal
$x$=0.5 and $x$=0.67 compositions and characterized by a
{\it ...-full-empty-..} and a {\it ...-full-full-empty-...} chain sequence
along the {\it a} direction, are considered
thermodynamically stable\cite{Schleger,Wille}. 
The superstructures arising from an ordering between oxygen-poor
and oxygen-rich chains are well described by a simple lattice gas
model, 
called ASYNNNI, first
introduced by de Fontaine {\it et al.}\cite{deFont}. Despite its
simplicity the ASYNNNI model, which considers only second nearest
neighbor interactions, can account for the stability region of the {\bf
OII} phase. Extensions of the model to include longer range
interactions\cite{Ceder} predict the occurrence of more complex superstructures
(e.g. the {\bf OIII} phase), even if they become significant only for very
well equilibrated samples, as it was systematically verified in the
range $0.67\le x\le0.75$\cite{Manca}.

 The understanding of oxygen
ordering in the Cu(1)O$_x$  plane and its effects on
superconductivity in 123 systems has been greatly enriched during
the last eleven years. 
It is by now clearly established that the charge transfer
process in 123 systems and the related superconducting properties
are a rather sensitive function not only of the oxygen content,
but also of the oxygen ordering in the Cu(1)O$_x$  plane
through its induced effects on hole density in the Cu(2)O$_2$
planes and consequently on $T_c$ \cite{Cava,Jorgensen}. The connection
between oxygen ordering in the chains and hole behavior in the
planes was already clearly manifested in the time dependent increase of
$T_c$ during room temperature annealing of samples produced by
fast quenching\cite{Jorgensen,VealA,VealB}. The formation of the {\bf OII}
superstructure is responsible for the 60K plateau
typically observed in the $T_c$ dependence from oxygen content in
123 and more recently the influence of {\bf OIII} 
 ordering on $T_c$ has been shown\cite{Calestani}.

The variety of possible
superstructures has raised the question of the existence, for each
different ordering scheme, of a characteristic $T_c$ of their
own\cite{Cava}. However, as stressed by Shaked et al.\cite{Shaked},
experiments that unambiguously prove this hypothesis are difficult
or impossible as a result of the difficulty of stabilizing an
entire sample in a particular ordered state and comparing such
sample with those having the same oxygen content but a different
ordering. The major limitation is the utilization of {\it single}
samples, prepared one at a time in conditions that make a
comparative study extremely difficult, owing to a lack in
reproducibility produced by the significant influence of
experimental conditions and  thermal history on the 123
properties. 

To investigate the effects of oxygen ordering we
recently proposed  a novel strategy based on oxygen-equalized
pair-samples\cite{Manca}, prepared simultaneously in the same thermal
conditions, one by intercalation and the other by deintercalation
of oxygen, the fully oxygenated and reduced ({\bf OI} 
and {\bf T}) end-terms of the YBa$_2$Cu$_3$O$_{6+x}$
system acting as oxygen donor and acceptor respectively, to arrive
at the final oxygen content $k$ in both samples. This
topotactic-like technique for low-temperature processing of
oxygen-equalized ($k$) deintercalated [D]$_k$ and intercalated
[I]$_k$ pair samples of YBa$_2$Cu$_3$O$_{6+x}$ allowed us to
investigate unanswered questions about the relationship between
structure and superconductivity in this system\cite{Manca,Calestani}. On the
basis of the acquired experience in controlling the process
reproducibility we are now able to explore, for the first time in
comparative way, the most important (and at the same time the most
difficult to study)  region in 123 system:  the transient {\bf T}-{\bf
O} boundary around $k$=0.30, characterized by the vanishing of
semiconducting antiferromagnetic (SAF) behavior and the emergence
of  superconductivity (SC). 

Bulk polycrystalline [D]$_k$ and
[I]$_k$ pair-samples, hereafter referred to as $k$-pairs, were
prepared in a reproducible way starting with fully oxygenated
{\bf OI} bar-shaped samples of (3.0x2.0x14.0)mm$^3$,
weighing each one about 0.5 g, prepared twenty at a time by
following conventional solid-state reactions and sintering, and
fully reduced {\bf T} samples obtained from the former
by dynamic vacuum annealing at 650 $^{\circ}$C. From iodometric
and weight-loss analyses, the quoted oxygen content in the
reference ({\bf OI}, $x = 0.96$) and in the derived
({\bf T}, $x = 0.07$) samples was estimated to be accurate
within 0.02 oxygen atom per formula unit. Individually weighed
{\bf OI} and {\bf T} bars were equilibrated
at a given temperature ($T_e$) and order-stabilized at
composition-dependent temperatures ($T_s$) within the thermal
stability domain of {\bf OII} and {\bf OIII} 
superstructures\cite{Yang,SchlegerB}. By varying the
{\bf OI}/{\bf T}  mass ratios it is possible
to prepare $k$-pairs
 in a wide range of equalized
oxygen stoichiometry $k$. The {\bf OI} ({\bf T})
mass loss (gain) is due solely to a change in oxygen content in
the Cu(1)O$_x$  plane\cite{Jorgen}, and excellent agreement
between calculated and experimental oxygen content at equilibrium
was systematically obtained. Details on starting materials and
$k$-pair processing were reported elsewhere\cite{Manca}. This
topotactic-like procedure yields pairs of 123 specimens under
equilibrium conditions with equal oxygen content and thermal
history.  The $k$-pairs under investigation ($0.28 \leq k \leq
0.32$) were obtained by thermal equilibration of {\bf OI} and 
{\bf T} samples at $T_e  = 670^{\circ}$ C for 1 day, slow cooling at 
0.2$^{\circ}$C/min to
$T_s =75 ^{\circ}$C followed by order-stabilization at this
temperature for 3 days and final cooling (0.2$^{\circ}$C/min) to
room temperature. Several batches were prepared in this way and
comparatively characterized by resistive ($\rho(T)$), electron
diffraction (ED)  and NQR studies. 

Displayed in fig.1 are the
evolution of the $\rho(T)$ curves (A) and the representative
densitometer traces of an ensemble of ED patterns (B) recorded
independently on several fragments of three typical samples.
Panels 1 in fig.1 show the transport (A$_1$) and structural
(B$_1$) characteristics of the $k$=0.28 samples. Resistivity is
thermally activated and only tetragonal peaks show up in the
diffraction pattern. Both [I] and [D] are therefore characteristic
of a non-superconducting tetragonal phase, for which $k$=0.28
defines the upper limit of existence in both samples. Panels 4
likewise show the corresponding lower limit ($k$=0.32) for the
existence of a  partially {\bf OII}-ordered
superconducting  phase. Note the diffuse $({1\over2}\,0\,0)$ peak in
diffraction patterns B$_4$ ( in agreement with the doubling of the
a axis  produced by a {\it ...-full-empty-...} chain sequence in the
Cu(1)O$_x$ plane) and the coincidence of T$_c$ in the [I]
and [D] curves A$_4$. The situation is totally different in $k=0.30$
pairs, which display a phase separation. Resistivity is shown in
panels A$_2$ and A$_3$. The [D] sample is insulating, but its
curve (A$_2$) displays a kink precisely at the same temperature
where the corresponding [I] sample shows (A$_3$ curve) the onset
of the SC transition, which percolates the bar. [D] grains
invariably show the two kinds of patterns displayed in panel B2:
most of them tetragonal (solid line) and a minority fraction
{\bf OII} (dotted line) characterized by very diffuse
spots. A similar separation is observed for the [I] samples: most
grains are characterized by diffuse {\bf OII} 
superstructure spots (solid line) or by diffuse (dotted line)
extra peaks at $({h\over3}\,0\,0)$, whereas only few are tetragonal. 

These
data indicate that the  {\bf T}-{\bf O} phase
transition in the $k=0.30$ pairs displays the coexistence
of tetragonal and orthorhombic domains. This result is consistent with the
prediction by de Fontaine {\it et al.}\cite{deFont} from a lattice-gas
model. However the systematic observation  of the $({h\over3},0\,0)$
spots adds a new detail to this picture. We believe that these spots 
result from domains of an orthorhombic anti-III
({\bf OIII}$^*$) structure characterized by an ideal
{\it ...-empty-empty-full-...} periodic arrangement of chains along the
{\it a} direction\cite{Reyes}. Such a sequence gives rise to a tripling of
the $a$ axis in analogy with the {\it ...-empty-full-full-...}
configuration for the ideal composition $x$=$2\over3$ of the {\bf OIII}
superstructure\cite{Stratilov,Schleger}. To our knowledge the
{\bf OIII}$^*$ structure is reproducibly observed around
the ideal stoichiometry $x$=$1\over3$ for the
first time by means of our equilibration
technique. 
After long-term aging (one year) of
[I]$_{0.30}$ samples at room temperature the $({h\over3}\,0\,0)$ spots
disappear, the original two-phase orthorhombic state
({\bf OII}+{\bf OIII}$^*$) stabilizes in the
{\bf OII}  single-phase state and the resistive SC
transition broadens considerably. Hence the {\bf OIII}$^*$
ordering appears to be a metastable precursor in the emergence of
{\bf OII} ordering in the [I]$_{0.30}$ SC samples. 
 
The {\bf OIII}$^*$ phase cannot be justified by the original ASYNNNI
model\cite{deFont}, 
due to the
neglect of long range interactions. These interactions were later 
introduced in an extended model\cite{Ceder}, 
limited to the $6.5\le x \le 7$ range, to account for the observation
of the {\bf OIII} phase. 
Our carefully equilibrated samples, which reproducibly develop both the 
{\bf OIII} ($k\sim$0.7)\cite{Manca} and the {\bf OIII}$^*$ ($k$=0.3) 
phase, call for an extension of the long range interaction 
models to the oxygen poor region of the phase diagram.

We investigated the local structure by NQR to determine the degree of
short range order in $k$=0.3 samples at the SAF-SC boundary
\cite{Sanna}. The NQR resonance frequency, proportional to the
electric field gradient (EFG) at the nucleus, is characteristic of
each distinct copper site in the lattice. Since there are two Cu
isotopes (63 and 65) each lattice site gives rise to an isotope
doublet, at fixed frequency ($\nu_{63}$/$\nu_{65}$=1.082 ) and
intensity (I$_{63}$/I$_{65}$=2.235) ratios. The Cu NQR spectra of
the pair [D]$_{0.30}$ and [I]$_{0.30}$ are plotted in fig. 2 in
the range 22-33 MHz, corrected for frequency dependent sensitivity
and relaxation. Each sample shows two isotope doublets, the solid
line being the best Gaussian fit to the above mentioned isotopic
constrains. The EFG values of the two doublets identify them as
two distinct Cu(1) sites:\cite{Lutgemeier} the 28.05-30.35 MHz doublet is
2-Cu(1), linearly coordinated with apical oxygen and neighbored by
oxygen vacancies (v-Cu$^{1+}$-v) in the plane, while the 22.1-23.9
MHz doublet corresponds to the chain-end configuration (O-Cu$^{2+}$-v)
of the 3-fold coordinated 3-Cu(1). The few 4-Cu(1) (O-Cu$^{2+}$-O) contribute
negligibly to the spectra because of their much larger EFG
inhomogeneity. 

The area $A_i$ ($i$=2,3)
under each doublet yields the average 
number of oxygen atoms in the inter-Cu(1) sites (i.e. the
average chain length) as
$\ell={3\over7}(2A_2/A_3+1)$, and we obtain $\ell_I=3.9(1)$
and $\ell_D=1.9(1)$ for the two samples of the pair. The short
average chain length  found in [D]$_{0.30}$ is consistent with its
broad NQR lines, since a short correlation length implies a broad
distribution of EFG values. These results outline the role of the
chain length in determining the chain hole-doping efficiency and
confirm directly the theoretical prediction by Uimin {\it et
al.}\cite{Uimin} that there is essentially no charge transfer from
chain fragments shorter than three oxygen atoms.  

The $k$-pair
method proves itself as an effective tool to extract more
detailed information (inaccessible to {\it single
sample} experiments) on the mechanism of short range oxygen-chain
ordering which characterizes the transient SAF-SC region. The
experimentally demonstrated inequivalence of the [D]$_{0.30}$ and
[I]$_{0.30}$ samples of a pair agrees
with previous analogous results 
obtained around the {\bf OII}-{\bf OIII} and
the {\bf OIII}-{\bf OI} transition
boundaries\cite{Manca,Calestani}. This leads us to conclude that different 
metastable
states exist near the thermodynamic equilibrium at a given
oxygen content and are connected with the vacancy ordering in the Cu(1)-O$_x$
chain system. 
Different kinetic and thermodynamic reaction paths
are realized during intercalation or deintercalation of oxygen and
result in inequivalent chain growth  processes revealed by ED
and NQR.  Moreover we point out that with our
equilibration scheme
structurally distinct domains occur in the same sample in the
transient region around $k$=0.3,
while the SC
transition in [I]$_{0.30}$ and the resistive kink in [D]$_{0.30}$
(fig.1 A$_2$-A$_3$) systematically occur at the same temperature. 
This suggests that a simultaneous electronic and structural phase 
separation takes place at the SAF-SC boundary,
where orthorhombic (SC) and tetragonal (SAF) domains coexist. They
originate nanoscopically and are critically dependent on
the chain growth process. We believe that the different chain
lengths observed in [I] and [D] samples  represents an 
experimentally determined
critical borderline between the vanishing of  SAF behavior and the
emergence of  SC in 123.

%
%
 \begin{figure}
 \caption{
Resistivity $vs.$ temperature (A) and representative densitometer traces
of ED patterns along the {\it a$^*$} direction of the
reciprocal lattice (B) for [D]$_k$ and [I]$_k$. Panel 1 and 4: data
from $k$=0.28 and $k$=0.32 pairs respectively; panel 2 and 3: 
[D] and [I] samples of a $k$=0.30 pair.}
 \label{fig1}
 \end{figure}
\begin{figure}
 \caption{
NQR spectra of sample [I]$_{0.30}$ and [D]$_{0.30}$ . The area under
each isotope doublet, indicated as 3-Cu(1) and 2-Cu(1) respectively, is
proportional to the number of Cu ions in that local environment.}
 \label{fig2}
 \end{figure}

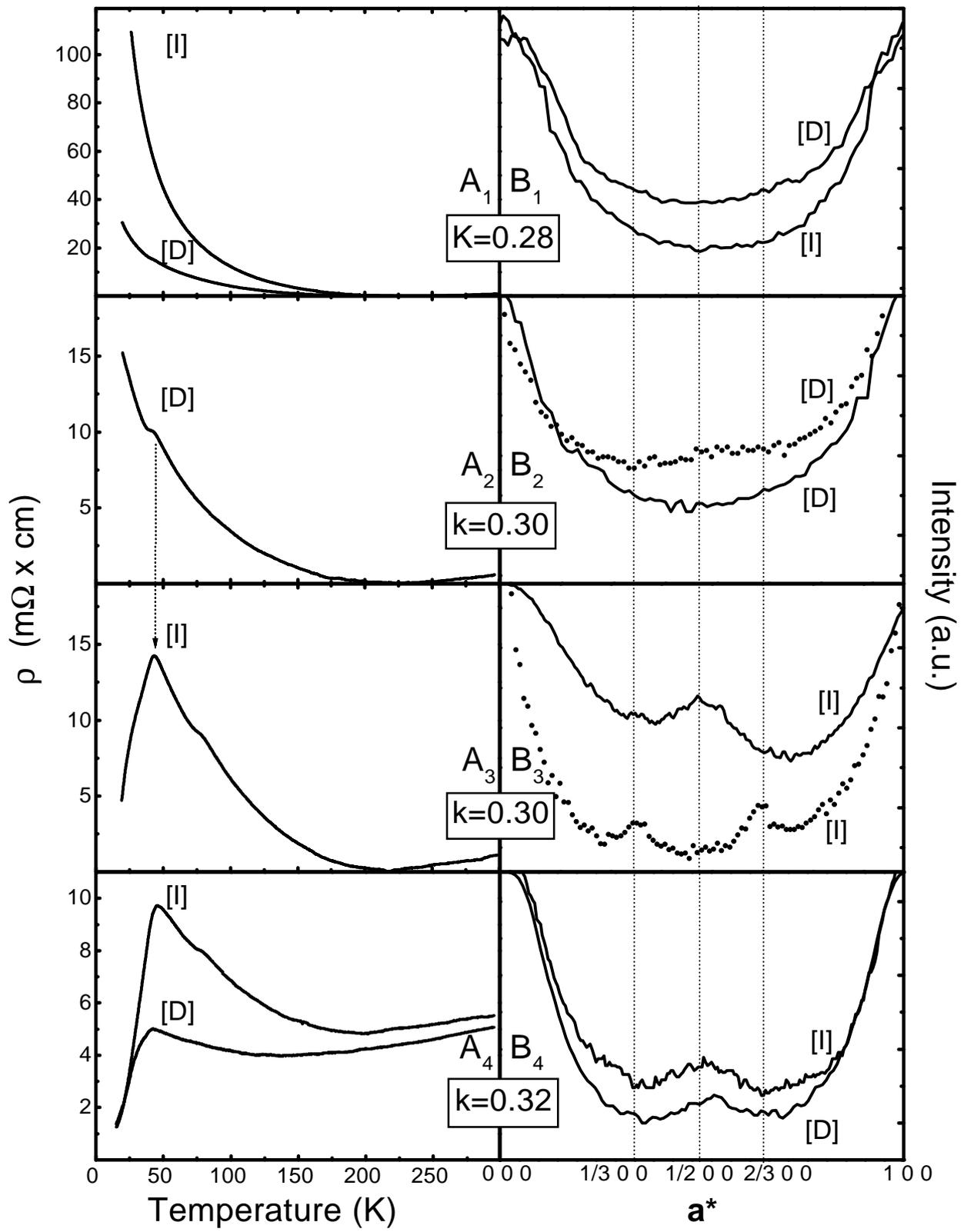
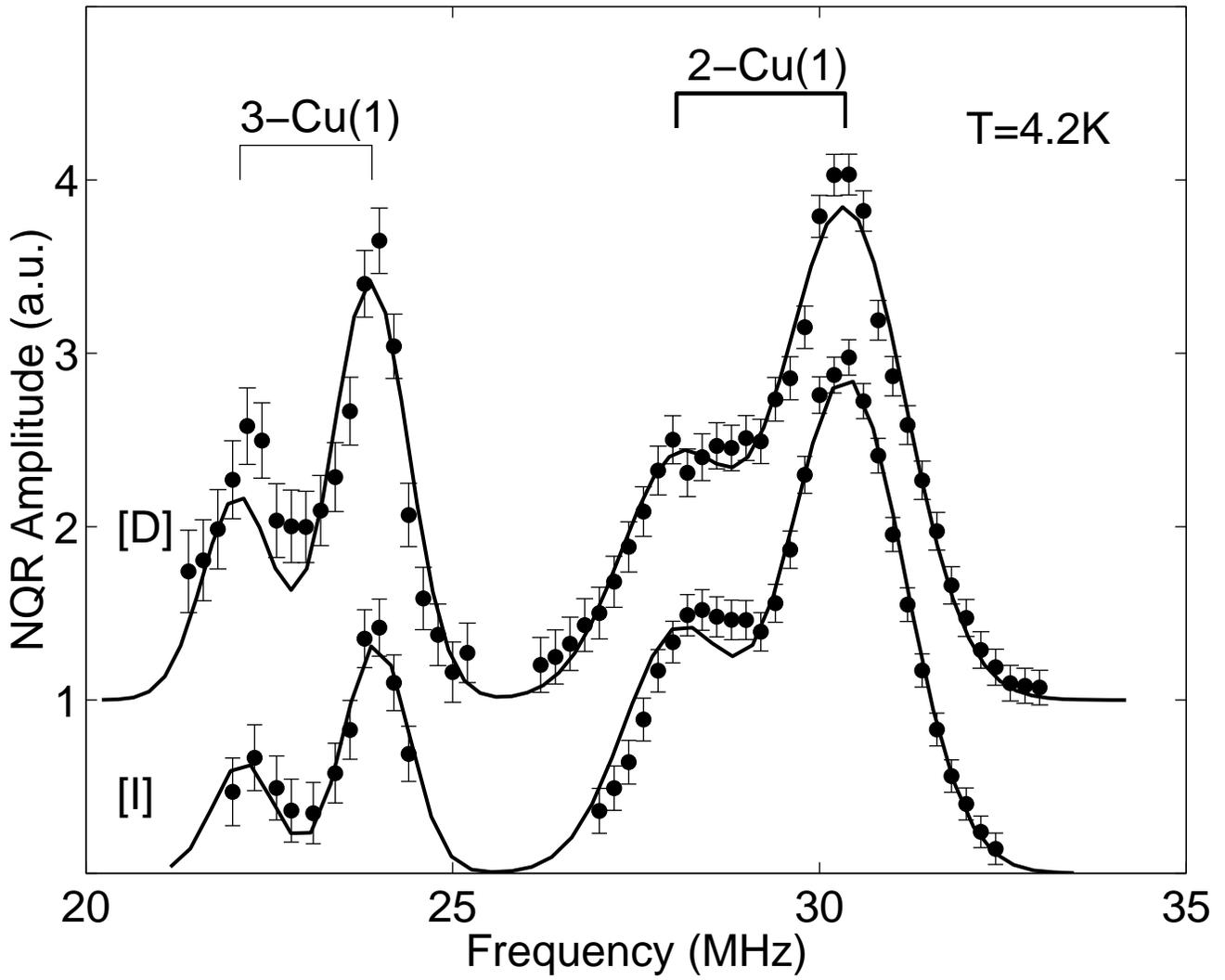
\end{document}